\documentclass[aps,prl,groupaddress,showpacs,twocolumn]{revtex4-1}
\usepackage{graphicx,bm,amsmath,amsfonts,amssymb,color,subfigure,bbold}

\newcommand{\be}{\begin{equation}}
\newcommand{\ee}{\end{equation}}
\newcommand{\bea}{\begin{eqnarray}}
\newcommand{\eea}{\end{eqnarray}}

\newcommand{\im}[1]{\mbox{Im}\left[#1\right]}
\newcommand{\re}[1]{\mbox{Re}\left[#1\right]}

\newcommand{\bx}{\bm{x}}

\newcommand{\pt}{\mathcal{PT}}
\newcommand{\apt}{\mathcal{APT}}
\newcommand{\pder}[2]{\frac{\partial #1}{\partial #2}}
\newcommand{\pderr}[2]{\frac{\partial^2 #1}{\partial #2^2}}
\begin{document}

\title{Antisymmetric $\pt$-photonic structures with balanced positive and negative index materials}

\author{Li Ge}
\email{lge@princeton.edu}
\author{H.~E.~T\"ureci}
\affiliation{Department of Electrical Engineering, Princeton University, Princeton, New Jersey 08544}

\begin{abstract}
We propose a new class of synthetic optical materials in which the refractive index satisfies $n(-\bx)=-n^*(\bx)$. We term such systems antisymmetric parity-time ($\apt$) structures. Unlike $\pt$-symmetric systems which require balanced gain and loss, i.e. $n(-\bx)=n^*(\bx)$, $\apt$ systems consist of balanced positive and negative index materials. Despite the seemingly $\pt$-symmetric optical potential $V(\bx)\equiv n(\bx)^2\omega^2/c^2$, $\apt$ systems are not invariant under combined $\pt$ operations due to the discontinuity of the spatial derivative of the wavefunction. We show that $\apt$ systems can display intriguing properties such as spontaneous phase transition of the scattering matrix, bidirectional invisibility, and a continuous lasing spectrum.

\end{abstract}

\maketitle

The pursuit of artificial structures exhibiting unusual electromagnetic properties is a major scientific endeavor. It has led to, for example, the development of photonic crystals for controlling the propagation of electromagnetic waves, utilizing band structures created by Bloch scattering in periodic structures \cite{PC}. Another achievement in this pursuit is the design of negative index materials (NIMs) aimed at subwavelength imaging \cite{pendry,Shalaev,Dolling,Valentine}. In NIMs the refractive index is negative over some frequency range, achieved by engineered electromagnetic resonances in nanostructures. Reduced intrinsic loss and even net gain have been demonstrated in NIMs by gain embedment \cite{Xiao}.

More recently, there has been a growing interest in systems that display parity-time ($\pt$) symmetry, both in quantum field theory \cite{Bender1,Bender2,Bender3} and optics \cite{El-Ganainy_optlett07,Makris_prl08,Regensburger}. By utilizing balanced gain and loss satisfying $n(-\bx)=n^*(\bx)$, many intriguing optical phenomena have been predicted and observed, such as double refraction \cite{Makris_prl08}, power oscillations \cite{Makris_prl08,ruter_natphy10,Zheng_pra10}, coexistence of coherent perfect absorption \cite{CPA, CPA_science} and lasing \cite{Longhi, CPALaser, Schomerus}, and unidirectional transmission resonances \cite{invisibility,PRA2012}.

In this Letter we propose a new class of synthetic systems bridging NIMs and $\pt$-symmetric photonics. Their refractive index is {\it antisymmetric} under combined $\pt$ operations, i.~e. $n(-\bx)=-n^*(\bx)$, with balanced positive index materials (PIMs) and NIMs. In addition, we require that the permeability satisfies $\mu(-\bx)=-\mu(\bx)$. The imaginary part of $n(\bx)$ is symmetric, which can be positive (loss), negative (gain), zero, or any complicated spatial function. We term such synthetic systems antisymmetric parity-time ($\apt$) structures, and we found that they can display intriguing features such as bidirectional invisibility, spontaneous phase transition of the scattering matrix, and a continuous lasing spectrum.

\begin{figure}
\centering
\includegraphics[width=0.9\linewidth]{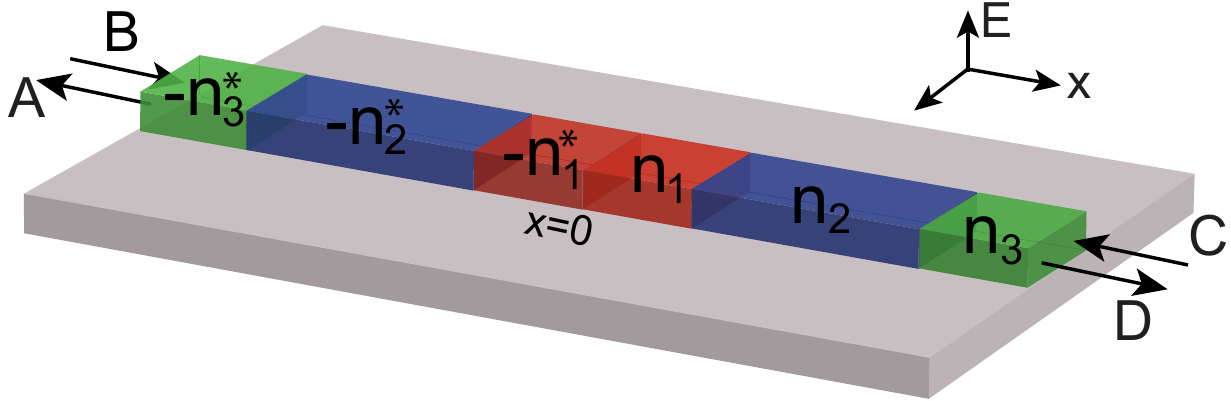}
\caption{(Color online) Schematic of a 1D $\apt$ photonic heterostructure, consisting of 6 layers with $n(-x)=-n^*(x)$.} \label{fig:1Dschematic}
\end{figure}

Below we base our discussion on the scalar wave equation for the electric field
\begin{equation}
  \left[\nabla^2 + n(\bx)^2\, (\omega^2/c^2)\right] E(\bx;\omega) = 0,
  \label{eq:helmholtz}
\end{equation}
which describes steady-state solutions for transverse electric waves in one-dimensional (1D) and two-dimensional (2D) systems. Henceforth we set $c=1$. By first glance one may think that all the phenomena found in $\pt$-symmetric systems would survive, since the optical potential $V(\bx)\equiv n(\bx)^2\omega^2$ is still invariant under $\pt$ operation. This is, however, not true due to the different boundary condition at PIM and NIM interfaces. Take a 1D $\apt$ heterostructure for example (see Fig.~\ref{fig:1Dschematic}), the electric field itself is still continuous at PIM and NIM interfaces, but its spatial derivation now satisfies \cite{NIM_cavityModes,NIM_scattering1D}
\be
\frac{1}{\mu_\text{PIM}}\left.\pder{E(x;\omega)}{x}\right|_{x\in \text{PIM}} = \frac{1}{\mu_\text{NIM}}\left.\pder{E(x;\omega)}{x}\right|_{x\in \text{NIM}}, \label{eq:BC_general}
\ee
which changes abruptly due to the sign difference of $\mu_\text{PIM}$ and $\mu_\text{NIM}$. Below We first analyze wave propagation and lasing in 1D $\apt$ heterostructures, followed by a discussion of pseudo-$\apt$ symmetry for wave propagation in 2D with the paraxial approximation.

The phase transition of the scattering matrix ($S$-matrix) in a $\pt$-symmetric system is predicted based on the invariance of the system under combined $\pt$ operations \cite{CPALaser}. We would not expect a similar phase transition in $\apt$ systems, since flipping the sign of the real part of the refractive index is not related to any symmetry operation of the physical state. However, there do exist some special properties of the transmission coefficient $t$ and the left and right reflection coefficients $r_L,\,r_R$ in a 1D $\apt$ heterostructure:
\bea
r_L = r_R^*, \quad \text{Im}[t]=0. \label{eq:r&t_1D}
\eea
To understand these properties, we start by noting one observation: By changing the refractive index of each layer in an {\it arbitrary} photonic hetereostructure to its negative complex conjugate and flipping the sign the magnetic pearmeability, i.e. $n\rightarrow -n^*,\,\mu\rightarrow-\mu$, the transfer matrix $M$, defined by
\be
\left( \begin{matrix} A\\ B\end{matrix}\right) = M
  \left(\begin{matrix} C\\ D\end{matrix}\right),
\ee
becomes its complex conjugate at the same {\it real} frequency:
\be
M(\omega) \rightarrow M^*(\omega),\quad \text{Im}[\omega]=0. \label{eq:M_APT}
\ee
The field amplitudes $A$, $B$, $C$, and $D$ are defined by
\be
E(x;\omega) = \left\{
\begin{array}{l l}
Ae^{-in_0\omega(x+L/2)} + Be^{in_0\omega(x+L/2)},& x<-L/2, \\
Ce^{-in_0\omega(x-L/2)} + De^{in_0\omega(x-L/2)},& x>L/2,
\end{array}
\right.
\ee
and illustrated in Fig.~\ref{fig:1Dschematic}. Here $n_0$ is the refractive index of the environment and we assume the corresponding $\mu_0=1$. The proof of (\ref{eq:M_APT}) is straightforward from the analytical expression of $M$:
\be
M(\omega) = D_0^{-1} \left[\Pi_{i=1}^N m_i\right] D_0,
\ee
obtained from the continuity of $E(x;\omega)$ and Eq.~(\ref{eq:BC_general}). The matrices $D_0$ and $m_i$ are given by
\begin{gather}
D_0 =
\begin{pmatrix}
1 & 1 \\
n_0 & -n_0
\end{pmatrix}, \\
m_i(\omega) =
\begin{pmatrix}
\cos(n_i\omega\Delta_i) & i\frac{\mu_i}{n_i}\sin(n_i\omega\Delta_i) \\
i\frac{n_i}{\mu_i}\sin(n_i\omega\Delta_i) & \cos(n_i\omega\Delta_i)
\end{pmatrix},
\end{gather}
where $\Delta_i$ is the width of the $ith$ layer. Under the transformation $n_i\rightarrow -n_i^*,\,\mu_i\rightarrow-\mu_i\,(i=1,\ldots,N)$, one finds $m_i(\omega)\rightarrow m_i^*(\omega)$ at a {\it real} frequency and so does $M(\omega)$. Since $M(\omega)$ determines the wave propagation, all related quantities such as $r_L$, $r_R$, and $t$, become their complex conjugate under this transformation.

Below we refer to the left half of an $\apt$ system $U$ and the right half $V$. Eq.~(\ref{eq:M_APT}) implies that the transfer matrices of $U$ and $V$ are related by
\be
M_V(\omega) = \sigma [M_U^{-1}(\omega)]^*\sigma, \quad \sigma=\begin{pmatrix} 0 & 1 \\ 1 & 0 \end{pmatrix}.
\ee
If we define $M_U(\omega)\equiv\left(\begin{smallmatrix} m_{11} & m_{12} \\ m_{21} & m_{22} \end{smallmatrix}\right)$ and the transfer matrix of the whole $\apt$ system $M_{\apt}(\omega)\equiv\left(\begin{smallmatrix} m'_{11} & m'_{12} \\ m'_{21} & m'_{22} \end{smallmatrix}\right)$, we then find through $M_{\apt}(\omega) = M_U(\omega)M_V(\omega)$ that
\begin{align}
&m'_{11} =  |m_{11}|^2-|m_{12}|^2, \nonumber \\
&m'_{22} = |m_{22}|^2-|m_{21}|^2, \label{eq:M_elements}\\
&m'_{12} = m_{12}m_{22}^*-m_{11}m_{21}^* = -(m'_{21})^* \nonumber.
\end{align}
The $S$-matrix defined by \begin{equation}
\left( \begin{matrix} A\\ D\end{matrix}\right) = S_{\apt} \left( \begin{matrix} B\\ C\end{matrix}\right) \equiv \left( \begin{matrix} r_L& t\\ t& r_R\end{matrix}\right) \left( \begin{matrix} B\\
C\end{matrix}\right) \label{eq:Sdef}
\end{equation}
can be expressed in terms of $M_{\apt}$:
\be
S_{\apt}(\omega) = \frac{1}{m'_{22}}\begin{pmatrix} m'_{12} & 1 \\ 1 & -m'_{21} \end{pmatrix},\label{eq:M2S}
\ee
from which we immediately find (\ref{eq:r&t_1D}) using (\ref{eq:M_elements}).

The phase transition of the $S$-matrix can be inferred from the relations (\ref{eq:r&t_1D}), which suggest the parametrization of the $S$-matrix by three independent {\it real} quantities: $t$, $a\equiv\text{Re}[r_L]$, and $b\equiv\text{Im}[r_L]$:
\be
S = \left(\begin{array}{c c}
        a+ib & t \\
        t & a-ib
       \end{array}\right). \label{eq:S}
\ee
We note that this general $S$-matrix is pseudo-Hermitian \cite{pseudoHermitian}, i.e. $S^\dagger=\eta S \eta^{-1}$ with $\eta = \left( \begin{smallmatrix}0&1\\1&0\end{smallmatrix}\right)$. The eigenvalues of the $S$-matrix are given by
\be
s_{\pm} = a \pm \sqrt{t^2-b^2},
\ee
which have two phases and the phase transitions occur at $t=\pm b$. The scattering eigenstates $\psi_\pm(\omega)=\left(\begin{smallmatrix} B_\pm \\ C_\pm\end{smallmatrix}\right)$ display a transition simultaneously:
\be
p_\pm \equiv \frac{B_\pm}{C_\pm} = \frac{1}{t}[ib\pm\sqrt{t^2-b^2}].
\ee

\begin{figure}
\centering
\includegraphics[width=0.7\linewidth]{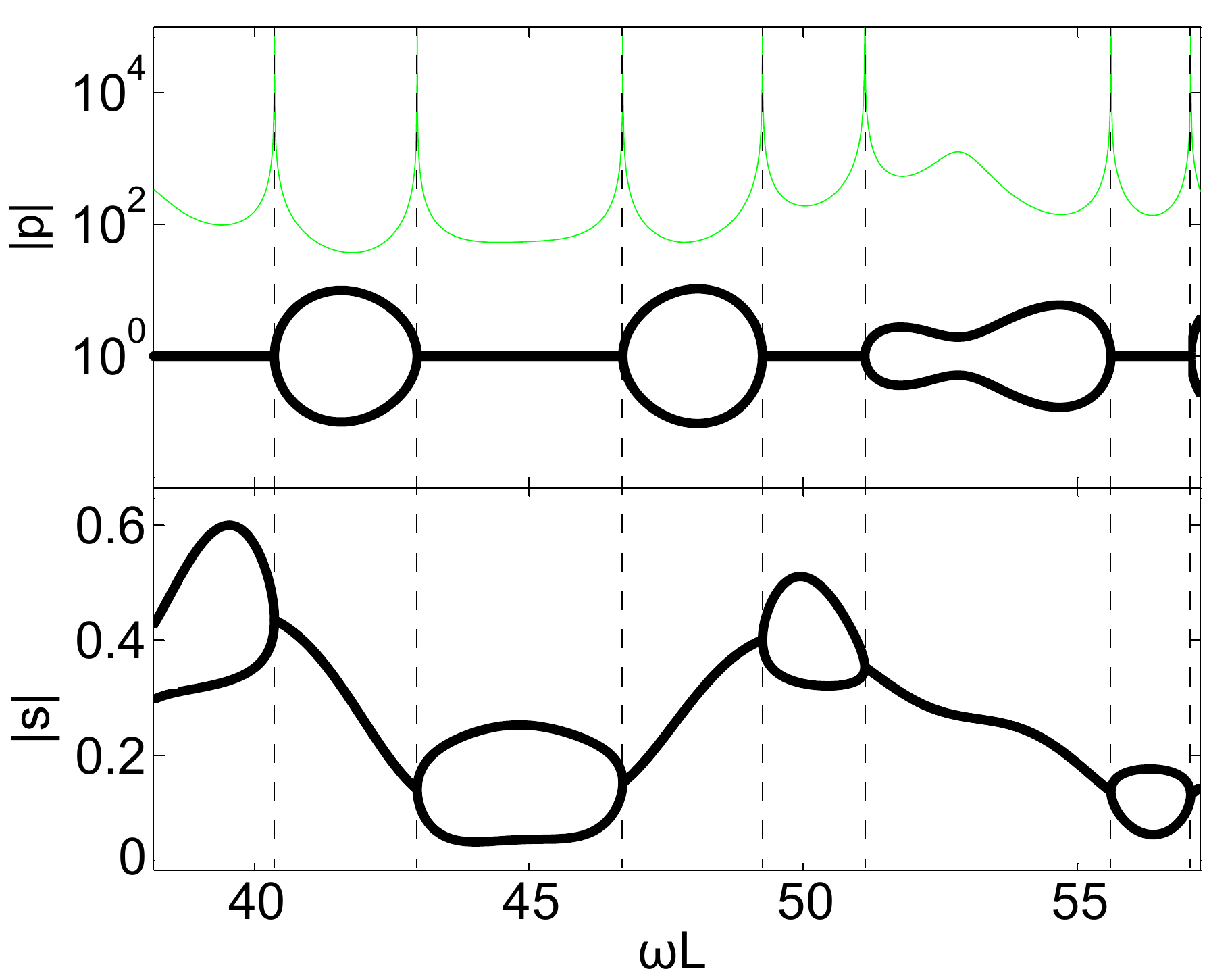}
\caption{(Color online) Phase transition of the $S$-matrix in an 8-layer $\apt$ heterostructure. For the four layers on the right ($x>0$), their widths are $\Delta_i=1.2, 0.996, 0.165, 0.531 {\mu}m$ ($i=1,2,3,4$), and their refractive indices are given by $\re{n_i}=1.3,-2,-1.7,-3$ and $\im{n(x)}=0.04$. Thick black solid lines in both panels show the transitions of the asymmetry factor $|p|$ and the scattering strength $|s|$. Thin green solid line in the upper panel indicates $|t^2-\text{Im}[r_L]^2|^{-1}$, which approaches infinity at the phase transition points (marked by dashed vertical lines).  } \label{fig:phaseTransition}
\end{figure}

In one phase ($|t|>|b|$) the intensities of the two incident beams are the same in a scattering eigenstate, i.e. $|p_\pm|=1$ (see Fig.~\ref{fig:phaseTransition}). Thus we refer to this phase as the symmetric scattering phase and the other as the symmetry-broken phase ($|t|<|b|$). In the symmetric phase $s_{\pm}$ are real, meaning that the symmetric inputs are either amplified or damped equally with no phase added during the scattering process.
%If $s_\pm<1$, then the two incident beams in a scattering eigenstate act as if they were reflected independently from the same real index material of $n_{eff,\pm} = n_0(1-s_\pm)/(1+s_\pm)$.
In the symmetry-broken phase the two scattering eigenstates have the same scattering strength $|s|$ ($s_+=s_-^*$) but with $|p_+| = |p_-|^{-1}$.
% and the phases of the two incident beams in a scattering eigenstate differ by $\pi/2$.

Tuning to the phase transition points requires adjusting either the gain and loss of the system, the frequency of the incident beams, or scaling the system size. Given that it is challenging to maintain $\apt$ (or $\pt$) symmetry in the first two approaches due to material dispersion, the third approach is probably most practical, i.e. by fabricating multiple scaled heterostructures and fixing the frequency of incident beams at a value that achieves the $\apt$ symmetry.

The phase transition discussed above is a general properties of all 1D $\apt$ systems, independent of whether the system has net gain or loss. By flipping the sign of $\im{n(x)}$, i.e. changing local gain into loss and vise versa, the system merely undergoes a time reversal and the phase transitions happen at exactly the same locations.

\begin{figure}
\centering
\includegraphics[width=0.7\linewidth]{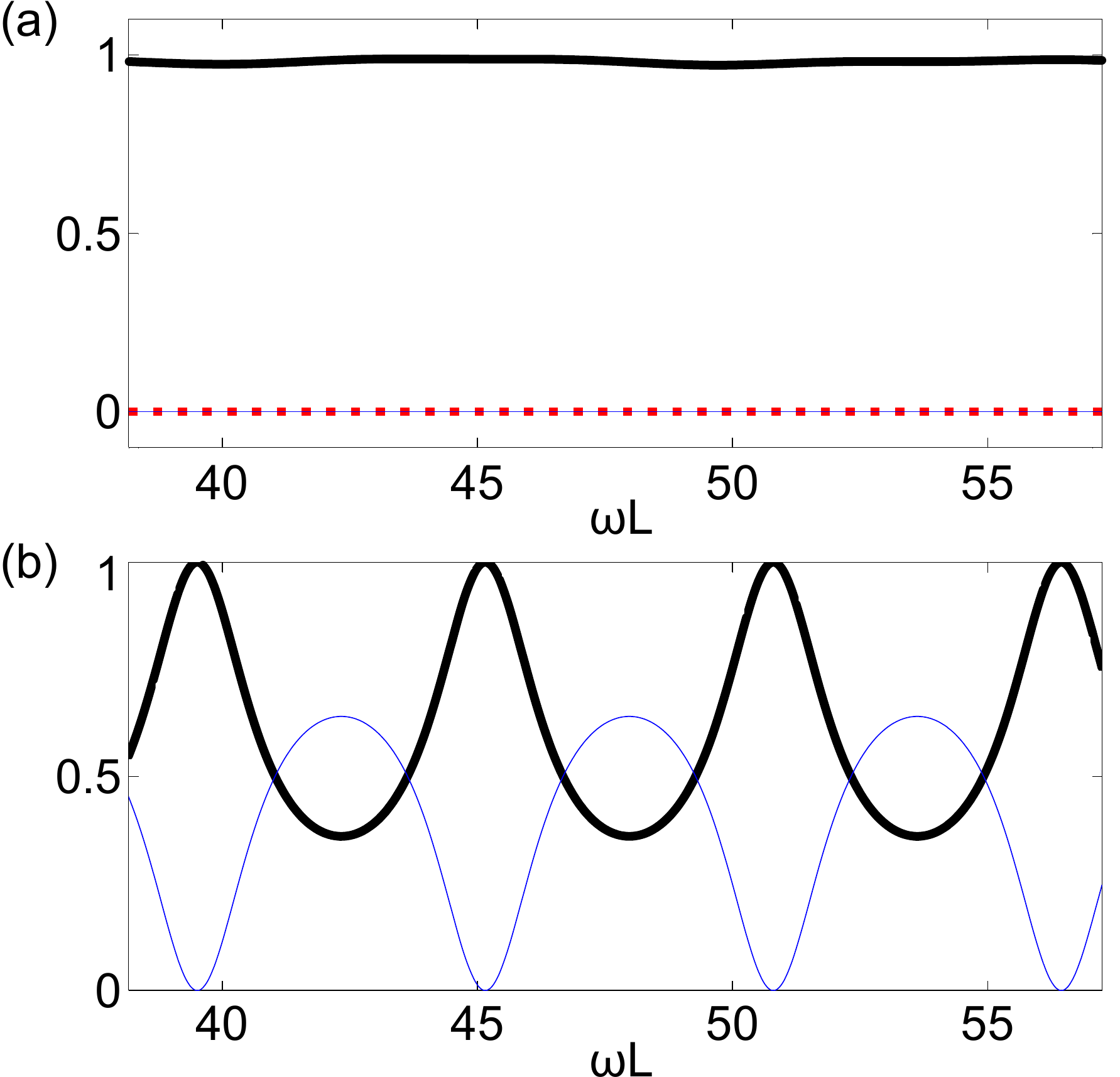}
\caption{(Color online) (a) Transmittance and reflectance in the $\apt$ photonic heterostructure studied in Fig.~\ref{fig:phaseTransition} but with $\im{n(x)}=10^{-4}$ (loss). Thick black line, thin blue line, and dotted red line indicate $T=|t|^2$, $R=|r_L|^2 \approx |r_R|^2$, and $\text{Arg}[t]$, respectively. They are very close to the values given by (\ref{eq:r&t_FTI}), i.e. 1, 0, 0, when $\im{n(x)}=0$. (b) Same as (a) but with $n_4=n_{-4}=3+0.04i$. The oscillations in $R$ and $T$ have a single period determined by the combined width of the leftmost and rightmost layers, i.e. $\Delta \omega L = {\pi L}/({2\Delta_4\re{n_{4}})}= 5.64$, as if the central 6-layer $\apt$ structure is absent. } \label{fig:1Dinvisibility}
\end{figure}

There is only one exception which occurs when the local gain/loss is zero \cite{gainInNIMs}, i.e. $\im{n(x)}=0$. In this case an $\apt$ heterostructure is always in the symmetric phase. More strikingly, the relations (\ref{eq:r&t_1D}) now become
\be
t=1,\quad r_L=r_R=0, \label{eq:r&t_FTI}
\ee
i.e. an $\apt$ system becomes invisible, and it is independent of the complexity and size of the heterostructure and at what frequency the $\apt$ symmetry occurs. This phenomenon is robust upon a slight breakdown of the $\apt$-symmetry or in the presence of a small $\im{n(x)}\neq0$ (see Fig.~\ref{fig:1Dinvisibility}(a)). In addition, the invisibility is independent of whether the $\apt$ structure is standalone or integrated in a photonic environment, as long as the neighboring elements are the same (see Fig.~\ref{fig:1Dinvisibility}(b)). If the $\apt$ symmetry can be maintained over a finite frequency range, a pulse transmitted within this frequency window will be exactly the same as the initial pulse, with no pulse distortion or shrinkage/expansion. This phenomenon is independent of the propagation direction, in contrast to the one-way invisibility found in $\pt$-symmetric heterostructures \cite{invisibility}.

The relations (\ref{eq:r&t_FTI}) can be treated as a generalization of vanished reflection that happens at the interface of two impedance matched PIM and NIM materials (see Ref.~\cite{pendry}, for example). There is at least one such interface in an $\apt$ system, i.e. at $x=0$, but multiple reflections occur at other interfaces between two NIMs, two PIMs, and a NIM and a PIM of different $|n|$. One way to prove (\ref{eq:r&t_FTI}) is from the transfer matrix $M_\apt(\omega)$ directly:
\be
M_\apt(\omega) = D_0^{-1} \left[\Pi_{i=-N}^{N} m_i\right] D_0, \quad i\neq0.
\ee
Note that
\be
m_{-i}(\omega)m_{i}(\omega) = \mathbb{1} \label{eq:pairCancelling}
\ee
when $n_i$ is real. Therefore,
\be
M_\apt(\omega) = D_0^{-1} \left[\Pi_{i=-N,\ldots,-2,2,\ldots,N}\, m_i\right] D_0 =\ldots= \mathbb{1}, \label{eq:M=1}
\ee
which implies (\ref{eq:r&t_FTI}).

\begin{figure}
\centering
\includegraphics[width=0.8\linewidth]{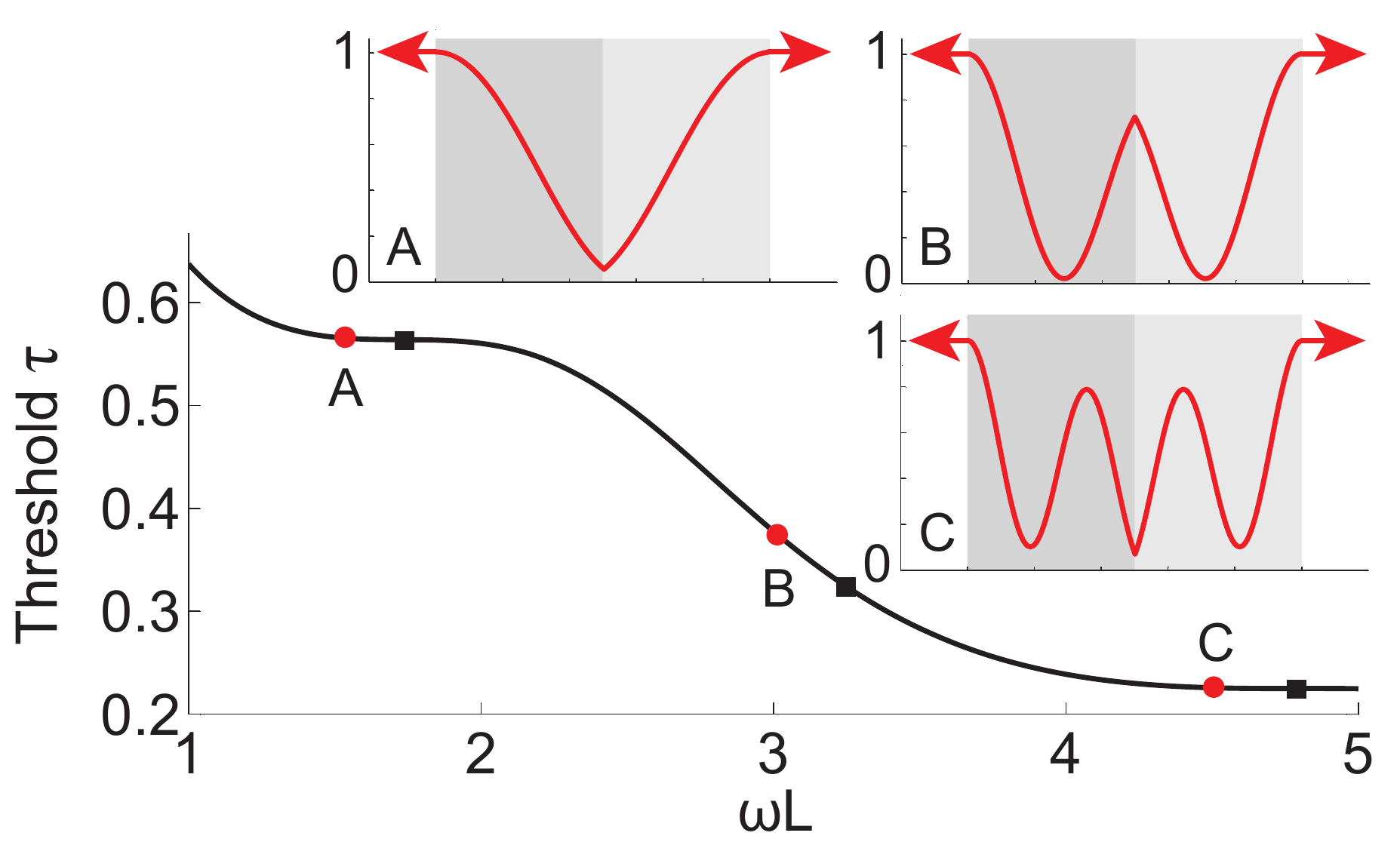}
\caption{(Color online) (a) Threshold value $\tau(\omega_L)$ for the continuous lasing spectrum in a 2-layer $\apt$ heterostructure. Each layer is 500nm thick and the refractive indices are $\pm2-i\tau(\omega_L)$ at threshold. Squares indicate the discrete lasing solutions $\omega L=1.735,\,3.245,\,4.786$ in a uniform PIM cavity of the same length. Inset: Representative examples of intensity profile in the continuous lasing spectrum. Their frequencies are chosen at $\omega L=1.5,\, 3,\,4.5$ and marked by the red dots in the main figure. Shadowed areas indicate the cavity.} \label{fig:lasing}
\end{figure}

Net gain has been demonstrated in NIMs by embedding an active medium \cite{Xiao}, which opens the possibility of achieving lasing in metamaterials. Lasing modes are given by the poles of the $S$-matrix (i.e. $m'_{22}=0$ in Eq.~(\ref{eq:M2S})) on the real frequency axis, which in general lead to a discrete set of solutions for the lasing frequency $\omega_L^{(m)}$ and the corresponding threshold $\tau^{(m)}=-\text{Im}[n(x)]>0$ (assuming a spatially uniform gain). Each lasing mode has its distinct intensity profile, and roughly speaking the mode order $m$ indicates the number of peaks inside the cavity. To determine $\{\omega_L,\tau\}$ of each mode in a 1D heterostructure, one can, for example, solve the two equations given by the real and imaginary part of $m'_{22}$. However, Eq.~(\ref{eq:M_elements}) shows that $m'_{22}$ is always real for an $\apt$ system, which implies that there exist continuous region(s) of $\omega_L$ in which a $\omega_L$-dependent $\tau$ can be found. In other words, an $\apt$ heterostructure has a continuous lasing spectrum and the mode order $m$ cannot be assigned. A 2-layer $\apt$ structure in the low frequency regime is shown as an example in Fig.~\ref{fig:lasing}. We observe a reduced threshold as the lasing frequency increases and more oscillations gradually appear in the intensity profile, which is symmetric since lasing occurs only in the symmetric phase of the $S$-matrix \cite{bibnote1}. If we consider the lasing modes in the corresponding PIM cavity of the same length and $|n|$, we find that its discrete lasing solutions lie exactly on the continuous threshold curve of the $\apt$ structure (see Fig.~\ref{fig:lasing} and Supplemental Material \cite{SI}). This comparison also shows that coherent feedback does occur in an $\apt$-system based laser, even though its spectrum is continuous.

Unlike the bidirectional invisibility, the continuous lasing spectrum is singular and breaks down if the $\apt$-symmetry is broken, which can be utilized as a sensitive measure of the quantities of interest that lead to the latter. Here we consider one scenario where the the $\apt$-symmetry is broken due to a slight length mismatch. Consider a 2-layer cavity of length $L$ similar to that studied in Fig.~\ref{fig:lasing} but with the PIM layer wider than the NIM layer by $\delta$. We found that all lasing modes disappear except the ones at $kL\approx m\pi L /(|n|\delta)\,(m=1,2,\ldots)$, whose thresholds are about the same as their values with $\apt$ symmetry. Take $L=250 \mu{m}$, $\delta = 250\, \text{nm}$, and $|n|=2$ for example, the corresponding wavelengths are $\lambda=m^{-1}\, {\mu}m$, which are are well separated and easy to achieve single-mode lasing. These modes originate from the resonances of the tiny section of length $\delta$, which acts as an external cavity for frequency selection. In this example, the variation of $\delta$ (or $L$) is enhanced by four times in the wavelength of the fundamental mode ($m=1$) since $\Delta\lambda = 4\Delta\delta/m$, which can be easily measured. As a comparison, the sensitivity to detect $\delta$ is reduced by a factor of $\delta/L=10^{-3}$ using a lasing mode of roughly the same wavelength in a uniform PIM cavity of length $L$, which also has a much denser spectrum to analyze. We note that the laser linewidths are comparable in the two systems, since the thresholds of the corresponding modes are about the same as discussed.

So far we have discussed the $\apt$ symmetry with balanced PIMs and NIMs. One may attempt to realize a ``pseudo-$\apt$'' symmetry using only PIMs (or NIMs), satisfying $n(-x)=-n^*(x)$ but with $\mu(-x) = \mu(x)$. Such a symmetry can be realized, for example, for wave propagation in 2D paraxial geometry \cite{El-Ganainy_optlett07, Makris_prl08}, with transverse index variation $n(x)=n_0+\delta n(x)$ satisfying $|\delta n(x)|\ll n_0$ and $\delta n(-x)=-\delta n^*(x)$. Eq.~(\ref{eq:helmholtz}) for the transverse electric field $E(\tilde{x},z) = \phi(\tilde{x},z)e^{ik_0z}$ propagating in the $z$ direction becomes \cite{Makris_prl08}
\be
i\pder{\phi(\tilde{x},z)}{z} + \left[\pderr{}{\tilde{x}} + \delta n(\tilde{x})k_0\right] \phi(\tilde{x},z) = 0, \label{eq:paraxial}
\ee
where $\tilde{x} \equiv \sqrt{2n_0k_0}x$ and $\phi(\tilde{x},z)$ is the slowly varying amplitude. The transverse optical potential now is proportional to $\delta n(\tilde{x})$ instead of $n^2$ in Eq.~(\ref{eq:helmholtz}), and the intriguing phenomena discussed above disappear. The only exception happens when the system becomes equivalent to a conventional $\pt$-symmetric structure. The latter occurs, for example, if $\delta n(\tilde{x})= A\sin\tilde{x}+iB\cos\tilde{x}\,(A,B\in\mathbb{R})$; shifting $\tilde{x}$ by $\pi/2$ transforms $\delta n(\tilde{x})$ to $A\cos\tilde{x}-iB\sin\tilde{x}$, satisfying $n(-\tilde{x})=n^*(\tilde{x})$.

In summary, we propose a new class of synthetic materials which are antisymmetric under combined parity-time operations, i.e.  $n(-\bx)=-n(\bx)^*$. $\apt$ systems demonstrate interesting features such as bidirectional invisibility, spontaneous phase transition of the $S$-matrix, and a continuous lasing spectrum. Properties of $\apt$ systems in higher dimensions are under investigation and will be reported elsewhere.

We acknowledge A. Douglas Stone and Kostas Makris for helpful discussions. This research was supported by MIRTHE NSF EEC-0540832.

\appendix
\section{Appendix: Threshold correspondence in a simple $\apt$ heterostructure and a PIM cavity}
\label{sec:threshold}
The threshold $\tau(\omega_L)$ of the continuous lasing spectrum in a simple 2-layer cavity discussed in the main text is given by the solution of the {\it real} equation
\begin{align}
&|\cos\alpha|^2 + \frac{n_r^2-\tau^2}{n_r^2+\tau^2}|\sin\alpha|^2 \nonumber \\
= &-\text{Im}\left[\left(n_r-i\tau + \frac{1}{n_r-i\tau}\right)\sin\alpha(\cos\alpha)^*\right], \label{eq:THcurve}
\end{align}
in which $n_r$ is the real part of the refractive index in the PIM, $\alpha\equiv(n_r-i\tau)\omega_LL/2$, and $L$ is the cavity length. We has set $c=1$ in the main text. In comparison, the threshold and the discrete lasing frequency in an uniform PIM cavity of the same length are simultaneously determined by the following {\it complex} equation
\be
\cos(2\alpha) = i\left[\left(n_r-i\tau + \frac{1}{n_r-i\tau}\right)\sin\alpha\cos\alpha\right]. \label{eq:TH1}
\ee
It implies that
\be
\tan\alpha = -i(n_r-i\tau),\,-\frac{i}{n_r-i\tau},
\ee
or
\be
\text{Re}\left[\frac{\tan\alpha}{(\tan\alpha)^*}\right] = \frac{n_r^2-\tau^2}{n_r^2+\tau^2}. \label{eq:TH2}
\ee
By taking the real part of both sides of (\ref{eq:TH1}) after multiplying them by $(\cos\alpha)^*/\cos\alpha$, we find
\begin{align}
&|\cos\alpha|^2 + \text{Re}\left[\frac{\tan\alpha}{(\tan\alpha)^*}\right]|\sin\alpha|^2 \nonumber \\
= &-\text{Im}\left[\left(n_r-i\tau + \frac{1}{n_r-i\tau}\right)\sin\alpha(\cos\alpha)^*\right],
\end{align}
from which we recover (\ref{eq:THcurve}) using (\ref{eq:TH2}).


\begin{thebibliography}{100}
\bibitem{gainInNIMs} An active medium is required in the NIM layers to compensate the intrinsic loss to reach this condition.
\bibitem{bibnote1}  Lasing occurs at the pole of the $S$-matrix, i.e. $m'_{22}=0$. Together with $\det(M)=1$, which holds for any arbitrary 1D structure with the same neighboring element on each side, we find $|m'_{12}|=|m'_{21}|=1$. This implies $|a\pm ib|\rightarrow|t|$ as we approach the pole condition, while $t=1/m'_{22}\rightarrow\infty$. Therefore, we find $|t|\geq|b|$, which indicates that the $S$-matrix is in the symmetric phase. Another way to see this is from the transfer matrix directly. Lasing requires $C=0,\, D\neq0$ (only outgoing waves), and $m'_{22}=0,\, |m'_{12}|=1$ lead to $|A|=|D|,\,B=0$.
\end{thebibliography}
\end{document}